\pdfoutput=1
\documentclass[acmsmall,screen]{acmart}

\usepackage{cleveref}
\usepackage{siunitx}
\usepackage{soul}
\usepackage{graphicx}
\usepackage{subcaption}
\usepackage[inline]{enumitem}

\usepackage[frozencache=true, cachedir=_minted-paper]{minted} 
\definecolor{bg}{rgb}{0.95,0.95,0.95}
\setminted[python]{breaklines}
\setminted[python]{xleftmargin=\parindent}
\setminted[python]{bgcolor=bg}

\newcommand{\code}[1]{\texttt{#1}}
\newcommand{\mypy}{mypy}

\newcommand{\Any}{\code{Any}}

\newcommand{\pct}[1]{\SI{#1}{\percent}}

\begin{document}

\title{Toward a Corpus Study of the Dynamic Gradual Type}

\author{Dibri Nsofor}
\orcid{0009-0003-7599-2657}
\affiliation{
  \department{Kahlert School of Computing}
  \institution{University of Utah}
  \city{Salt Lake City}
  \state{Utah}
  \postcode{84112}
  \country{USA}
}
\email{dibrinsofor@gmail.com}

\author{Ben Greenman}
\orcid{0000-0001-7078-9287}
\affiliation{
  \department{Kahlert School of Computing}
  \institution{University of Utah}
  \city{Salt Lake City}
  \state{Utah}
  \postcode{84112}
  \country{USA}
}
\email{benjamin.l.greenman@gmail.com}

\begin{abstract}
  Gradually-typed languages feature a dynamic type that supports
  implicit coercions, greatly weakening the type system but making types easier to
  adopt.
  Understanding how developers use this dynamic type is a critical question
  for the design of useful and usable type systems.
  This paper reports on an in-progress corpus study of the dynamic type in
  Python, targeting 221 GitHub projects that use the mypy type checker.
  The study reveals eight patterns-of-use for the dynamic type,
  which have implications for future refinements of the mypy type
  system and for tool support to encourage precise type annotations.
\end{abstract}

%

\maketitle


\section{Introduction}
\label{s:intro}
\label{s:background}

Programming languages have traditionally forced developers to choose
between the ``flexibility and simplicity'' of untyped code~\cite{h-hints-1989}
and the maintenance benefits of static types~(e.g.~\cite{h-icfp-keynote}).
Gradual typing offers a compromise through a flexible \emph{dynamic type} that
can be used in any context without a type error~\cite{st-sfp-2006,svcb-snapl-2015}.
For example, a function that expects an argument of the dynamic type (named
\Any{} in Python) can treat the argument as though it has any precise type;
the code below assumes its dynamic argument is a class, adds a method to it,
and returns the updated class:

\begin{minted}{python}
import types; from typing import Any # the dynamic type in Python

def addPrice(cls: Any) -> Any: # mixin, add method to class
  cls.price = types.MethodType(lambda self: 99, cls)
  return cls
\end{minted}

\noindent%
A caller can send any value to the \code{addPrice} function without raising
a compile time type error.
Incorrect calls, such as \code{addPrice(11)} result in a runtime error when executed.

The dynamic type adds a degree of optimism to the typechecker, fundamentally
changing the nature of type analysis.
Without the dynamic type, a well-typed program is certifiably made of parts
that fit together~(it cannot ``go wrong''~\cite{m-jcss-1978}).
With the dynamic type, a well-typed program is code that \emph{may} fit together provided that 
every occurrence of the dynamic type receives well-behaved values at runtime---which, as noted 
in work on static blame~\cite{sclz-jfp-2024}, may be impossible.
Moreover, the dynamic type renders type analysis unsound
(unless coercions are backed by runtime checks, see~\cref{s:mypy}):
any value can inhabit any type after coercion.
The dynamic type also
pokes loopholes into other traditional properties.
For example, the simply-typed lambda calculus is strongly normalizing,
but loses this property with
the addition of a dynamic type~\cite{mf-toplas-2009}.

Given the perils of the dynamic type, it ought to be used sparingly: either as
a temporary measure in prototype software, or as a last resort when precise types
do not exist.
And yet, its use is widespread.
In our collection of 221 projects that use the mypy typechecker~(\cref{s:approach}),
there are 28,478 occurrences of the dynamic \Any{} type.
Understanding why these \Any{}s appear in widely-used code is a critical
question for the adoption of types.
If more-precise alternatives exist,
that motivates tool support for writing types.
If the \Any{} types cover up limitations of the typechecker, that motivates research
on type system design.
If the rationale for certain \Any{} types is unclear, that motivates
interviews and developer surveys.

This paper is a progress report of our journey to discover the precise reason behind
\Any{} types across a large corpus of Python projects.
Based on a combination of manual and script-driven analysis on a small number of sample
projects, it presents:

\begin{itemize}
  \item
    the design of a corpus study to study use of the dynamic type~(\cref{s:approach}),
  \item
    eight usage patterns~(\cref{s:pattens}) and discussions of how to identify the patterns automatically.
\end{itemize}

A major goal of this paper is to solicit early-stage feedback on the design
of our corpus study experiment.
Automated software analysis is a powerful but dangerous tool, as flaws in the
protocol can lead to questionable
conclusions~\cite{rpfd-fse-2014,bhmvv-toplas-2019,bmvv-arxiv-2019,rdf-arxiv-2019}.
In particular, our methods for detecting usage patterns must guard against
false positives before we apply them at scale.
The paper concludes with
future work~(\cref{s:next}),
related work~(\cref{s:rw}), and a discussion that positions this study
in the broader context of gradual languages~(\cref{s:conclusion}).

\section{Background: Optional Typing and Mypy}
\label{s:mypy}

Gradually-typed languages come in many varieties because the question of how to enforce static types 
in the midst of untyped code exposes a complex design space.
Full enforcement via behavioral contracts
is a compelling vision~\cite{tf-dls-2006,ff-icfp-2002}, but leads to high run-time 
overhead~\cite{gtnffvf-jfp-2019}
without custom runtime support~\cite{fgsfs-oopsla-2018,bbst-oopsla-2017,kas-pldi-2019}.
One alternative is to enforce only the top-level shape of types, but this can
still lead to overheads of 2x or more~\cite{vss-popl-2017,g-pldi-2022,gm-pepm-2018}.
A second alternative is to forbid untyped data structures from entering typed
code~\cite{wzlov-popl-2010,lgmvpk-pj-2023,mt-oopsla-2017}.
A third alternative is to forgo enforcement (and type soundness) altogether.
This third option is the most popular choice~\cite{gdf-toplas-2023}, and is
known as {optional typing}~\cite{b-pluggable-2004}.

Mypy is an optional typing system for Python~\cite{mypy}.
It equips the PEP~484 syntax for types~\cite{pep484}
with static checks that detect inconsistencies between types and code.
Mypy can efficiently analyze millions of lines of code~\cite{dropbox-mypy}
and has been widely adopted.
Mypy supports annotations on variable, class, and function declarations.
Developers can annotate as many or as few of these positions as they choose.
By default, unannotated positions receive the dynamic \Any{} type, though
developers can change this default through {configuration options}
that fine-tune mypy or by writing a comment of the form
\code{\#type:ignore[ERR]}
with \code{ERR} replaced by a specific mypy error code~\cite{mypy-errcode}.
For programs that depend on untyped libraries, developers can write interfaces,
called \emph{type stubs}, that declare types for library exports.
The (untyped) definitions of these exports are not typechecked, but their uses (in
typed code) are typechecked.

In summary, there are several ways that the \Any{} type can enter a mypy codebase:
explicitly through developer-provided \Any{} annotations,
implicitly through configuration options,
and explicitly through lines with \code{\#type:ignore} comments.
All three sources are important for a corpus study.
But whereas configuration options and comments come with metadata to describe
how they systematically incorporate an \Any{} type, developer-provided
annotations require manual analysis.
The main focus of this paper is on the manual analysis.

\section{Corpus Study Design}
\label{s:approach}

\begin{figure}[t]
  \centering
  {\includegraphics[width=0.66\columnwidth,trim={140 120 140 100}]{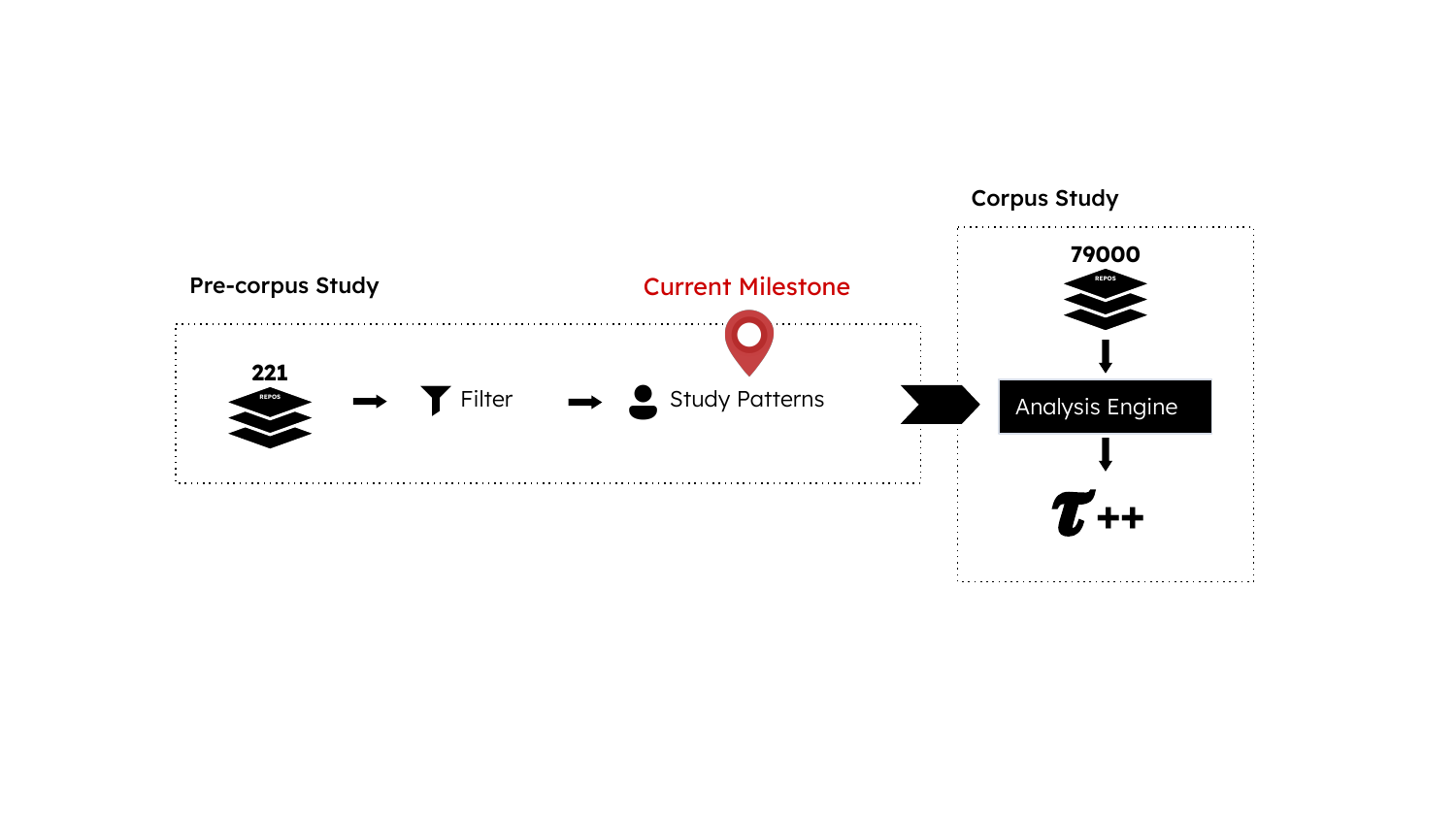}}
  \caption{Project overview with a marker at the current milestone: Study Patterns}
  \label{f:approach}
\end{figure}

Our goal is to learn why developers resort to the dynamic type
by analyzing thousands of open-source projects.
Currently, we are in the midst of an initial, {formative} stage
of the project (herein called the pre-corpus study), in which we study instances of the dynamic
type by hand.
As patterns emerge, the next step is to build automated tools and measure
how common each pattern is across our corpus of approximately 79,000 projects.
\Cref{f:approach} sketches this overall pipeline and highlights the current focus.
This section describes our protocols for building a corpus~(\cref{s:approach:1}),
filtering uninteresting instances of the \Any{} type~(\cref{s:approach:2}), and
manually finding patterns~(\cref{s:approach:3}).

\subsection{Building the Corpus}
\label{s:approach:1}

We built a corpus by querying GitHub via Mozilla's
\code{agithub} utility~\cite{agithub} and filtering the results.
The query asked for Python projects with a modest number of GitHub stars (80 stars)
and at least one of the following configuration files, which may contain
\mypy{} options:
\code{mypy.ini}, \code{pyproject.toml}, \code{setup.cfg}, and \code{config}.
This query resulted in over 79K projects.
We chose a sample of 221 projects that had at least 250 GitHub stars and
that actually used \mypy{} (confirmed manually)
as a target for the pre-corpus study.

\subsection{Filtering Any Types}
\label{s:approach:2}

Mypy comes with a utility called \code{stubgen} that extracts the type annotations
from a Python file into
a type stub interface~(\cref{s:mypy}).
We use these stubs as the first step toward manual analysis.
The stubs from our 221 sample projects contain 318,564 lines of annotations that include
the \Any{} type.

Many of the annotations are of low priority for building formative hypotheses.
First, method annotations whose only \Any{} is in the first argument position
are common and often uninteresting; these come about because
\mypy{} infers \Any{} when a method argument lacks an annotation.
Second, a file may contain several duplicate signatures.
After filtering both sorts of annotations, we arrived
at 41,447 distinct lines in which to search for patterns.

The majority of annotations in project files are for functions or methods.
We further classified these arrow types into three categories using a script:
methods with \Any{} for their first parameter (i.e., \code{self} parameter),
arrows with \Any{} inside a callback argument (\code{Callable[T1,T2]}),
and arrows with \Any{} inside a dictionary argument (\code{Dict[T1,T2]} or
\code{Mapping[T1,T2]}).
These categories helped to guide our manual analysis.

\subsection{Manually Studying Patterns}
\label{s:approach:3}

Both authors of this paper manually examined type annotations and code to identify patterns of use
for the \Any{} type.
One author studied every annotation in 10 large projects for a total of over 5K annotations.
The other author took an ad-hoc sample of 30 projects (spread throughout the alphabet,
from \code{BentoML} to \code{wandb}) and sampled types in those projects that
used \Any{} inside of a \code{Callable} or \code{Dict} type, inspecting a total
of 92 annotations.
Whenever the authors found an \Any{} type that seemed likely to reveal typing challenges,
they searched for its origin in the source code to learn more.



\section{Patterns of Dynamic Type Use}
\label{s:pattens}

Our manual study of types and code in a few dozen sample projects has revealed eight
ways in which developers appear to use the \Any{} type in mypy (implicitly and explicitly).
These eight categories range from specific anti-patterns to vague cases
where precise types exist but the code uses \Any{} nevertheless.
Below, we present each pattern with a code example and discuss how source-code
analysis might identify the pattern at scale.
For some patterns, we have implemented a related analysis; where applicable, we report
data based on our 221 sample projects.

\subsection{Dynamic Instead of a Type Variable}
\label{s:pat:tvar}

Every occurrence of the \Any{} type is a wildcard.
By contrast, type variables can
introduce a degree of uniformity.
Consider this equality function from our corpus:

\begin{minted}{python}
def eq(a: Any, b: Any) -> bool:
  return a == b
\end{minted}

\noindent%
Every call of the form \code{eq(x, y)} is type correct, even though
the function cannot return \code{True} when its arguments have different types.
If the type signature used a variable in place of the two \Any{}s,
developers would know more about the function's behavior
(even though mypy cannot easily report a type error, see~\cref{s:pat:uvar}).
If the intent is indeed to allow type-mismatched inputs, the type could
say so with distinct variables for its two parameters. 

We do not have an automatic analysis for this pattern.
It is unclear what to look for in a type signature:
although type variables would help this \code{eq} function,
other functions that take two \Any{}s as input may not benefit.
As another example, we have found {reduce} functions that return an \Any{}
and therefore fail to describe type constraints between their inputs and outputs.
But, not every function with \Any{} as its return type is parametrically polymorphic
in the same way as a reducer.

\subsection{Implicit Dynamic Due to Unconstrained Type Variables}
\label{s:pat:uvar}

Two occurrences of the same type variable must be instantiated with
matching types.
Mypy's default notion of matching is rather loose because of subtyping, and thus may lead to surprising outcomes.
In the following example, the call to \mintinline{python}{count_cars} instantiates
the variable \mintinline{python}{Car} as both a string and a number; this
is not a type error because strings and numbers have a common supertype in mypy:

\vspace{-1ex}
\begin{minted}{python}
Car = TypeVar('Car') # car is unconstrained / unbounded
Traffic = Union[Car, List['Traffic']]

def count_cars(x: Traffic, car: Car) -> int:
  if isinstance(x, List):
    x.append(car)
  return len(x)

count_cars(["FJ40", "Baja Buggy"], 5) # NOT a type error
\end{minted}

\noindent%
To prevent such weakening, type variables must be declared with either
a set of constraints
or an upper bound~(e.g.,
 \mintinline{python}{TypeVar('Car', str, bytes)}
 or \mintinline{python}{TypeVar('Car', bound=AnyStr)}).

Unconstrained type variables are straightforward to detect: there are
471 of them across our 221 sample projects.
Discovering where these variables lead to weak types is another matter,
and this calls for instrumentation within mypy to report variables that get 
instantiated with a top type.

\subsection{Dynamic Instead of a Self Type}
\label{s:pat:self}

Some methods that return their receiver object use \Any{} as the return type to
allow for subclass polymorphism.
In the example below, the return type
\code{Shape} would be too conservative
(it would prevent the use of subclass methods after \code{Circle().move(...)})
and the return type \code{Circle} would be unsound.
Using \Any{} gets rid of spurious errors by skipping type analysis altogether:

\begin{minted}{python}
class Shape:
  def move(self, dist: int) -> Any: # imprecise return type, Self is better
    self.position += dist
    return self

class Circle(Shape):
  pass

Circle().move(4)
\end{minted}

\noindent%
A precise alternative is \code{Self} as a return type, which propagates
the type of the receiver to the result.
Mypy added support for \code{Self} in version 1.0~\cite{mypy-v1};
code in our corpus evidently needs an upgrade, either because of
a knowledge gap or because it was written before the \mypy{} 1.0 release.
To flag this pattern, an analysis must find methods that return their first parameter
and use \Any{} as the return type.


\subsection{Dynamic for Dependent Dictionaries}
\label{s:pat:ddict}

The mypy type \mintinline{python}{Dict[T1, T2]} describes simple dictionaries
with keys of type \code{T1} and values of type
\code{T2}.
This type cannot express dictionaries in which specific keys point to values of different
types.
In the example below, for instance, the key \code{"price"} assumes float
values:

\begin{minted}{python}
def get_discount(item: Dict[str, Any]) -> int:
  if "price" in item: 
    discount = item["price"] * 0.15
    return item["price"] - discount
\end{minted}

This is a modest form of dependent types that appears throughout
our corpus.
In the 221 sample projects, there are 6,831 signatures that use
\Any{} for dictionary values
(i.e., \code{Dict}, \code{Dict[T1,Any]}, or \code{Mapping[T1,Any]}).
Row polymorphism using scoped labels~\cite{d-tfp-2005}
or singleton types (as in Haskell~\cite{ew-haskell-2012},
Typed Racket~\cite{tf-popl-2008}, or Elixir~\cite{cdv-pj-2024})
may suffice to improve mypy.
Alternatives are to adapt TypeScript's indexed access types~\cite{tskeyof},
introduce a metaprogramming layer for type tailoring~\cite{wcfg-ecoop-2024},
or integrate an SMT solver~\cite{v-dissertation-2016,kkt-pldi-2016}.

\subsection{Dynamic for Method Overrides}
\label{s:pat:override}

TypeScript infamously allows covariant method overrides---even though they are
unsound---because developers apparently want to write subclass methods that assume a narrow
set of inputs~\cite{bat-ecoop-2014}.
Mypy prevents covariant overrides by default, but developers can opt for unsound
overrides by writing a special comment after the subclass method signature.
Such comments do not use \Any{} directly, but are nevertheless relevant to our
study of how developers choose to weaken a type checker.
The following example is from the \href{https://github.com/ivre/ivre}{\code{ivre}}~\cite{ivre} project:



\begin{minted}{python}
class BinaryIO(IO[bytes]):
  def write(self, s: Union[bytes, bytearray]) -> int:
    pass

class FileOpener(BinaryIO):
  def write(self, s: bytes) -> int: # type: ignore[override]
    return self.fdesc.write(s)
\end{minted}


We have written an analysis for this pattern that finds these special comments,
looks for the parent class (which may be in a different file, or inaccessible to us in another project),
and examines the relationship between the parent and child method types.
There were 652 occurrences of this special comment across our sample corpus.
We found the parent class for 146 such comments, and \pct{80.8} (118 / 146) of these cases
use different argument types for parent and child methods.
These may be covariant overrides, but further analysis is needed to confirm.



\subsection{Dynamic for Wrapper Functions}
\label{s:pat:wrapper}

A wrapper takes a function as input and returns a modified or enhanced
version of the function.
The wrapper below, for instance, takes a function \code{fn} that expects a
statistics object and returns a function that expects a string; the wrapped
function converts input strings to objects and calls the original \code{fn}
function:

\begin{minted}{python}
def validate_stat(fn: Callable) -> Callable:
  def string_fn(self, stat: str, *args, **kwargs) -> Callable:
    stat = string_to_stat(stat)
    return fn(self, stat, *args, **kwargs)
  return string_fn
\end{minted}

\noindent%
The type for this wrapper is merely \code{Callable -> Callable}, which
says nothing about the close relationship between the input and output
functions.
It may be possible to adapt types for variable-arity
polymorphism~\cite{stf-esop-2009} to directly express the concept of wrappers.

We have written an analysis for this pattern.
It examines functions that have a parameter with the plain type \code{Callable}
and reports a match if this parameter is called always with
catch-all positional and keyword arguments.
There are 906 such functions across the sample projects.

\subsection{Dynamic to Hide Unnecessary Details}
\label{s:pat:details}

When types require details that are unimportant to the code at hand,
developers can use the \Any{} type to fill in the holes.
For example, this function from the
\href{https://github.com/agateau/nanonote}{\code{nanonote}}~\cite{nanonote} project
simply returns the last value from a dictionary.
Any dictionary is a valid input, so the signature asks for a value
of type \code{Dict[Any, Any]} (a type variable would be reasonable here):

\begin{minted}{python}
def _get_dict_last_added_item(dct: Dict[Any, Any]) -> Any:
  return list(dct.values())[-1]
\end{minted}

Similar examples come from parameterized types such as
\code{PathLike[T]} or \code{Callable[T1, T2]}.
Replacing the parameter with \Any{} is a simple, valid choice.
It might be more precise to ask for a top parameterized type, but that requires
thought about whether each parameter should use the top type (\code{object}) or
bottom type (\code{Never}) depending on whether it is co- or contra-variant.

We do not have an automatic analysis for this pattern.
One idea is to collect all uses of a parameter such
as \code{dct} and see what constraints the uses
impose on its type parameters.
If there are no constraints and the parameter type is \Any{}, we have a match.

\subsection{Uncategorized Dynamic Types}
\label{s:pat:lazy}

Many \Any{} occurrences do not fit a clear pattern.
Some could easily be replaced
with precise alternatives.
For example, the \href{https://github.com/wandb/wandb}{\code{wandb}}~\cite{wandb} project
contains a function that always raises an exception but gives \code{Any}
rather than \code{Never} (or \code{NoReturn}) as its return type:

\begin{minted}{python}
def __getattr__(self, key: str) -> Any:
  if not key.startswith("_"):
    raise wandb.Error(f"...")
  else:
    raise AttributeError
\end{minted}

Refining such types may not be a priority for developers.
Perhaps they would accept a patch generated through the use of a typing tool,
such as Scotty~\cite{hfs-oopsla-2022} or MonkeyType~\cite{monkeytype}.






\section{Future Work}
\label{s:next}

Our work is the first step toward a large-scale corpus study to discover how
the dynamic type is used in practice.
We have identified eight patterns
and we have plans to implement analyses for five of
them~(\S\S\ref{s:pat:uvar}, \ref{s:pat:self}, \ref{s:pat:override}, \ref{s:pat:wrapper}, \ref{s:pat:details}).
Next, we need to implement these analyses and develop methods for
the remaining patterns: type variables~(\S\ref{s:pat:tvar}),
dependent dictionaries~(\S\ref{s:pat:ddict}),
and some uncategorized cases~(\S\ref{s:pat:lazy}).
Along the way, we may of course discover additional patterns
or refine the current patterns into more-specific categories.

In parallel, we can work to improve mypy.
Dependent dictionaries~(\S\ref{s:pat:ddict})
and wrapper functions~(\S\ref{s:pat:wrapper})
call for a more expressive typechecker.
The patterns related to type variables~(\S\S\ref{s:pat:tvar}, \ref{s:pat:uvar}),
self types~(\S\ref{s:pat:self}),
and omitted details~(\S\S\ref{s:pat:details},\ref{s:pat:lazy})
would benefit from tools that suggest type improvements.
In fact, implementing a tool may be the most
effective way to gain insight about uncategorized \Any{}s.

\section{Related Work}
\label{s:rw}

There have been several corpus studies related to gradual types.
\citet{rmmhd-dls-2020} study types in 2,678 Python projects from GitHub.
Those projects contained
thousands of instances of the \Any{} type, which reflects our experience.
Their study focuses on which types get used and whether the types actually typecheck,
whereas our work investigates the reasoning behind the \Any{} type to suggest
type system improvements.
\citet{jzdfl-ase-2021} study the use of non-\Any{} types in Python projects
to uncover design patterns for complex types, providing insights to grow the
type system in a way that is complementary to our work.
They find that \Any{} is the second most common type, behind \code{Optional}, with over
12K instances across the revision histories of 19 projects.
\citet{lhwp-saner-2023} apply multiple typecheckers to 13 Python projects to compare
the defect reports; they do not study the adoption of types.
Corpus studies of types in R~\cite{tgkv-oopsla-2020,gv-oopsla-2019}
and Ruby~\cite{dsf-plateau-2009} provided further inspiration for our work.

Other works have studied types and code quality.
\citet{bm-msr-2022} study the effect of TypeScript types on quality metrics, such as the number of
code smells and the bugfix-commit ratio, with an eye toward the \Any{} type.
They find that \Any{} correlates positively with code smells, which suggests that removing \Any{}
is valuable, but they also find no correlation to bugfix commits, which may mean that \Any{}
is not an impediment to maintenance.
\citet{kcvm-ieee-2022} examine 400 bugs that had been committed to 210 Python repositories
and attempt to detect these bugs by adding mypy annotations.
\citet{gbb-icse-2017} perform a similar experiment with 400 bugs in 398 JavaScript projects,
adding TypeScript and Flow annotations.
Both studies concluded that at least \pct{15} of the bugs could have been caught with types.
\citet{xcsglzx-icpc-2023} conduct a similar study using 40 bugs across 10 Python projects
and find that the use of three typecheckers can detect \pct{35} of bugs with no additional
annotations.
Adding types improves the detection rate to \pct{72}.



We remark that corpus studies alone cannot tell the whole story about the use of the dynamic type.
Code in public repositories says nothing about the trial-and-error development
work that led to the latest checkpoint.
Our study is even more limited because it considers only one version of each codebase,
unlike some related works~\cite{bm-msr-2022,kcvm-ieee-2022,gbb-icse-2017,xcsglzx-icpc-2023}.
Other types of studies are needed to capture the bigger picture.
Interviews and observations of developers at work
(such as~\cite{gcdph-icse-2024,bjp-oopsla-2023})
would give highly-detailed data on sample developers.
Surveys are a lightweight alternative that can reach a broader audience~\cite{bask-icer-2018,ab-sigcse-2015,bgimgm-cse-2016,sjw-jfp-2018,blpgz-icse-2016,lkft-icfp-2023,tgpk-dls-2018}.
Developers might submit examples of how and why they use the dynamic type to an
online or IDE-integrated form.
Telemetry that automatically detects and reports usage of the dynamic
type is a third option.
Privacy is a concern with telemetry;
by way of mitigation,
prior works have relied on summary statistics~\cite{gsnk-pj-2024,transparent-telemetry}
and differential privacy~\cite{zhlbr-oopsla-2020,hlzbr-ecoop-2021}.

A final point to consider is whether lab studies can effectively assess use of
the dynamic type.
Lab studies are typically confined to small, self-contained tasks
(as in~\cite{r-cs-1981,cmh-icse-2022,rhg-icse-2023,cams-oopsla-2020,rswhbgkmfs-spe-2023}),
but the benefits and challenges of the dynamic type shine most in complex projects.
With the dynamic type, code has more flexibility in the face of evolution
and offers less guidance to developers trying to learn the codebase.
How to effectively study these tradeoffs in a lab setting is an open challenge.

\section{Discussion}
\label{s:conclusion}
\label{s:usecases}

The dynamic type is widely used, and therefore appears crucial for gradual type
adoption.
It also weakens type analysis, and is thus best as a temporary stop along the
road to precise types.
We have identified eight patterns of dynamic
type use in mypy through a combination of manual (in dozens of projects)
and automated analysis (in 221 projects).
While it remains to be seen how often the patterns arise in a wider
sample and thus how critical each pattern is at large,
the results so far suggest improvements to mypy,
such as dependent types for configuration dictionaries~(\cref{s:pat:ddict}),
and to the mypy ecosystem, such as a tool to insert \code{Self} types~(\cref{s:pat:self}).

Looking beyond mypy, the long-term goal of this work is to develop a method for
gradual type system design:
start by extending a conventional type system with the dynamic type,
observe how developers use the dynamic type,
and introduce precise types that meet developers where they are.
We expect that several patterns we have found, such as dynamic instead of a type
variable~(\cref{s:pat:tvar}), will lead to useful observations in related languages
including
Luau~\cite{luau-hatra-2021,bfj-hatra-2023} and
TypeScript~\cite{bat-ecoop-2014}.
Such patterns belong in a toolkit that delivers insights from
\emph{big code}~\cite{vy-ftpl-2016} to find common friction points and thereby
balance type system expressiveness with usability.

\begin{acks}
  Thanks to Ashton Wiersdorf and Amanda Funai for their comments on early drafts.
\end{acks}

\newpage

\bibliographystyle{ACM-Reference-Format}
\bibliography{bib.bib}

\end{document}